\documentclass[11pt]{article}
\usepackage{moriond,epsfig}

\def\bea{\begin{eqnarray}}
\def\eea{\end{eqnarray}}

\newcommand{\be}{\begin{equation}}
\newcommand{\ba}{\begin{eqnarray}}
\newcommand{\ee}{\end{equation}}
\newcommand{\ea}{\end{eqnarray}}  

\def\lesssim{\mathrel{\hbox{\rlap{\hbox{\lower4pt\hbox{$\sim$}}}\hbox{$<$}}}}
\def\gtrsim{\mathrel{\hbox{\rlap{\hbox{\lower4pt\hbox{$\sim$}}}\hbox{$>$}}}}

\def\gtsima{$\; \buildrel > \over \sim \;$}
\def\ltsima{$\; \buildrel < \over \sim \;$}
\def\gsim{\lower.5ex\hbox{\gtsima}}
\def\lsim{\lower.5ex\hbox{\ltsima}}
\def\simgt{\lower.5ex\hbox{\gtsima}}
\def\simlt{\lower.5ex\hbox{\ltsima}}
\def\simpr{\lower.5ex\hbox{\prosima}}

\def\simless{\mathbin{\lower 3pt\hbox
   {$\rlap{\raise 5pt\hbox{$\char'074$}}\mathchar''7218$}}}   
\def\simgreat{\mathbin{\lower 3pt\hbox
   {$\rlap{\raise 5pt\hbox{$\char'076$}}\mathchar''7218$}}}   

\begin{document}
\vspace*{4cm}
\title{COSMIC STRUCTURE FORMATION AT HIGH REDSHIFT}

\author{ILIAN T. ILIEV}
\address{Astronomy Centre, Department of Physics \& Astronomy, 
Pevensey II Building, University of Sussex, Falmer, Brighton BN1 9QH, 
United Kingdom}
\author{KYUNGJIN AHN}
\address{Department of Earth Science Education, Chosun University, 
Gwangju 501-759, Korea}
\author{JUN KODA, PAUL R. SHAPIRO}
\address{Department of Astronomy, University of Texas, Austin, 
TX 78712-1083, U.S.A.}
\author{UE-LI PEN}
\address{Canadian Institute for Theoretical Astrophysics, University
         of Toronto, 60 St. George Street, Toronto, ON M5S 3H8, Canada}

\maketitle\abstracts{
We present some preliminary results from a series of extremely large, 
high-resolution N-body simulations of the formation of early nonlinear 
structures. We find that the high-z halo mass function is inconsistent
with the Sheth-Tormen mass function, which tends to over-estimate the 
abundance of rare halos. This discrepancy is in rough agreement with 
previous results based on smaller simulations. We also show that the
number density of minihaloes is correlated with local matter density,
albeit with a significant scatter that increases with redshift, as 
minihaloes become increasingly rare. The average correlation 
is in rough agreement with a simple analytical extended Press-Schechter 
model, but can differ by up to factor of 2 in some regimes.}

\section{Introduction}
The properties of the first luminous objects in the universe remain a big
enigma at present due to the scarcity of observational data. These objects, 
the first stars and galaxies, started forming very early, only about 100-200 
million years after the Big Bang, and their ionizing radiation eventually
completely reionized the intergalactic medium \cite{2006PhR...433..181F}. 
This complex process can be studied through numerical 
simulations 
and semi-analytical models
. The former have the advantage of being able to describe complex situations 
and, in particular, the non-linearities
of the cosmic structures, but are expensive to run and have limited dynamic 
range. The latter are much cheaper to run and thus allow studies of the full
parameter space, but inevitably involve many approximations and simplifications.   
The distribution and properties of the low-mass halos which host the first stars 
is an important ingredient in any semi-analytical model of reionization. 
Such small-scale structure potentially has very different properties from
larger structures we see at later times, as they probe a very different 
part of the initial power spectrum of density fluctuations. Therefore, it 
is important to check the validity of any models and fits to the halo mass 
function and bias in this new regime.

\section{Simulations}
The results we present in this work are based on series of very large 
N-body simulations, as summarized in Table~\ref{summary_N-body_table}.   
They follow between $1728^3$ (5.2 billion) and $5488^3$ (165 billion)
particles (the latter at present is the largest cosmic structure 
formation simulation ever performed) in a wide range of box sizes from 
$2/h$~Mpc up to $3.2/h$~Gpc. Spatial resolution ranges from 50 pc/h to
40 kpc/h, while the particle masses range from 100~$M_\odot$ up to 
$\sim6\times10^{10}$, which yields minimum resolved halo masses (with 20 
particles) between $2\times10^3M_\odot$ and $10^{12}M_\odot$. The simulations 
were performed with the code 
CubeP$^3$M\footnote{http://www.cita.utoronto.ca/mediawiki/index.php/CubePM}
\cite{2008arXiv0806.2887I}.

\begin{table*}[ht]
\vspace{-5mm}
\caption{N-body simulation parameters. Background cosmology is 
based on the WMAP 5-year results.}
\label{summary_N-body_table}
\begin{center}
\begin{tabular}{@{}|llllll|}
\hline
boxsize & $N_{part}$   & mesh   & spatial resolution & $m_{particle}$ & $M_{halo,min}$ 
\\[2mm]\hline
2$\,h^{-1}$Mpc & $2048^3$ & $4096^3$ & $48.8\,{h^{-1}}$pc & $99.8\,M_\odot$ & $1996\,M_\odot$
\\[2mm]
6.3$\,h^{-1}$Mpc & $1728^3$ & $3456^3$ & $182\,{h^{-1}}$pc & $5.19\times10^3\,M_\odot$ & $1.04\times10^5\,M_\odot$
\\[2mm]
11.4$\,h^{-1}$Mpc & $3072^3$ & $6114^3$ & $186\,{h^{-1}}$pc & $5.47\times10^3\,M_\odot$ & $1.10\times10^5\,M_\odot$
\\[2mm]
20$\,h^{-1}$Mpc & $5488^3$ & $10976^3$ & $182\,{h^{-1}}$pc & $5.19\times10^3\,M_\odot$ & $1.04\times10^5\,M_\odot$
\\[2mm]
114$\,h^{-1}$Mpc & $3072^3$ & $6144^3$ & $1.86\,{h^{-1}}$kpc & $5.47\times10^6\,M_\odot$ & $1.09\times10^8\,M_\odot$
\\[2mm]
1$\,h^{-1}$Gpc & $3072^3$ & $6144^3$ & $16.3\,{h^{-1}}$kpc & $3.62\times10^9\,M_\odot$ & $7.24\times10^{10}\,M_\odot$
\\[2mm]
3.2$\,h^{-1}$Gpc & $4000^3$ & $8000^3$ & $40.0\,{h^{-1}}$kpc & $5.67\times10^{10}\,M_\odot$ & $1.14\times10^{12}\,M_\odot$\\
\hline 
\end{tabular}
\end{center}
\end{table*}

\begin{figure}[ht]
\begin{center}  
\includegraphics[width=3.in]{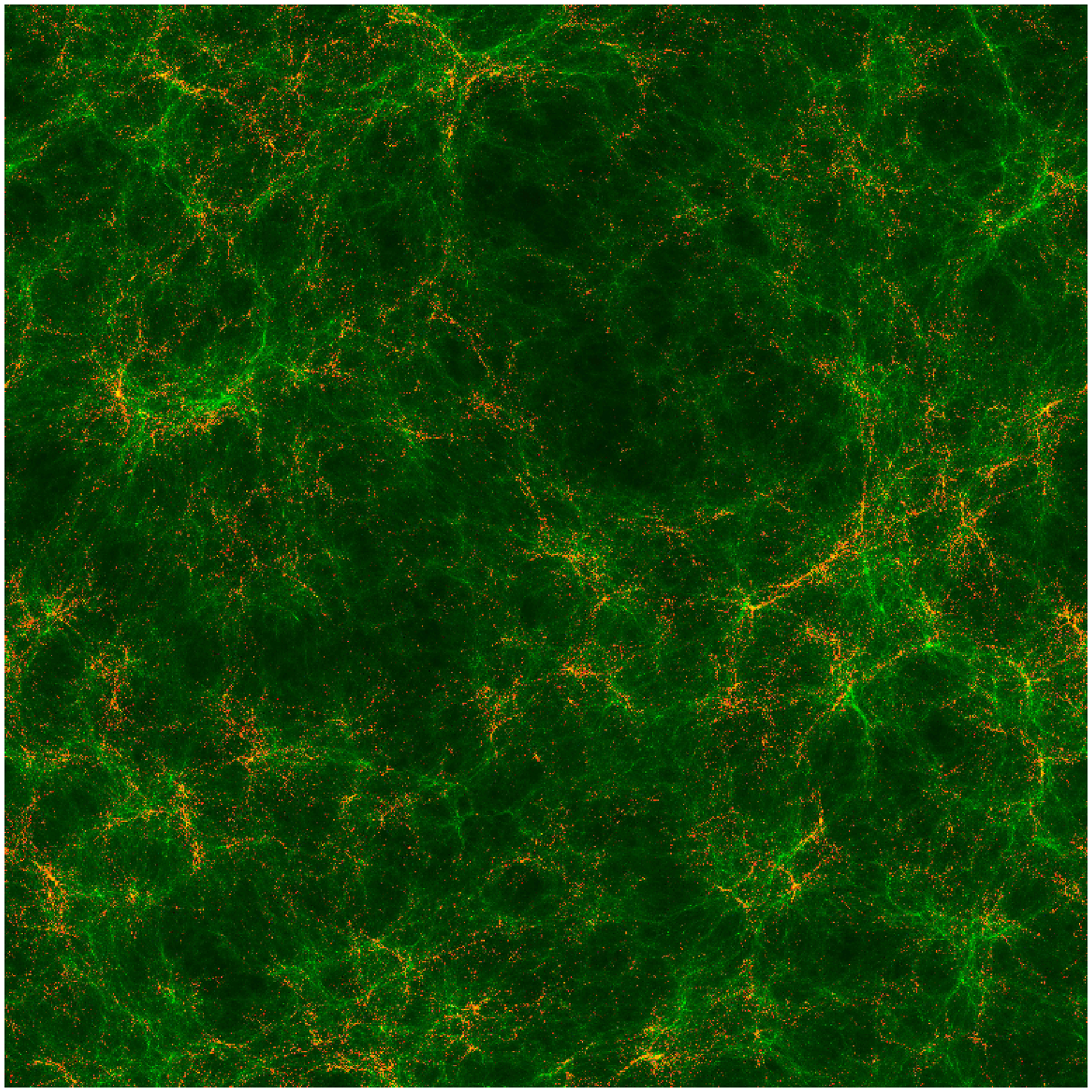}
\includegraphics[width=3.in]{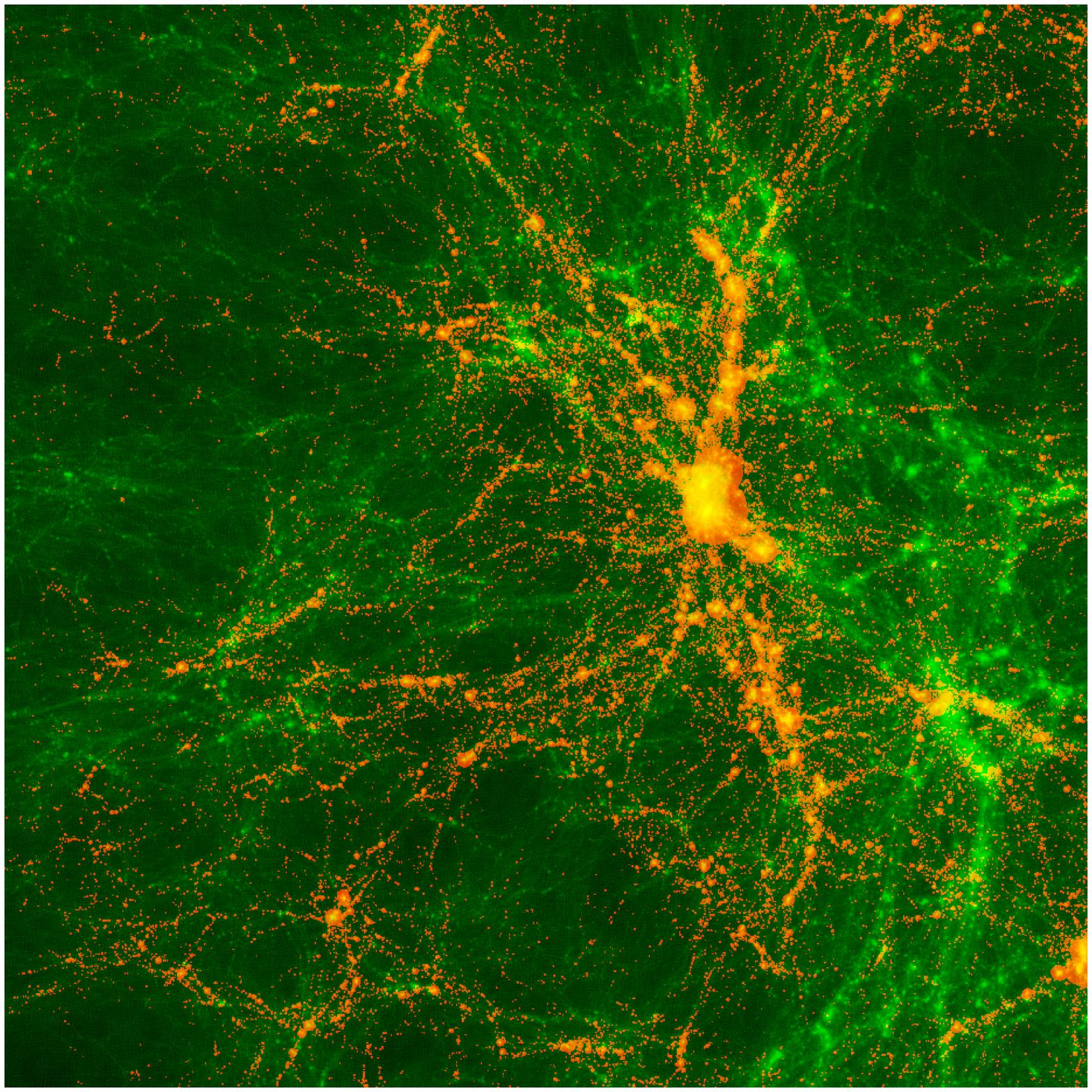}
\caption{Cosmic Web at redshift $z=8$ from simulation with boxsize 
$20{\,h^{-1}}$~Mpc and $5488^3=165$ billion particles resolving halos with 
minimum masses of $10^5M_\odot$. Shown are projections of the total 
density (green) and halos (actual size; orange). Slice is 444~$\,{h^{-1}}$kpc 
thick, images are of the full box (left) and of a zoomed sub-region 
1.8$\times$1.8~$\,{h^{-1}}$Mpc in size (right).
\label{z9_density_image_fig}}
\end{center}
\begin{center}  
\includegraphics[width=2.in]{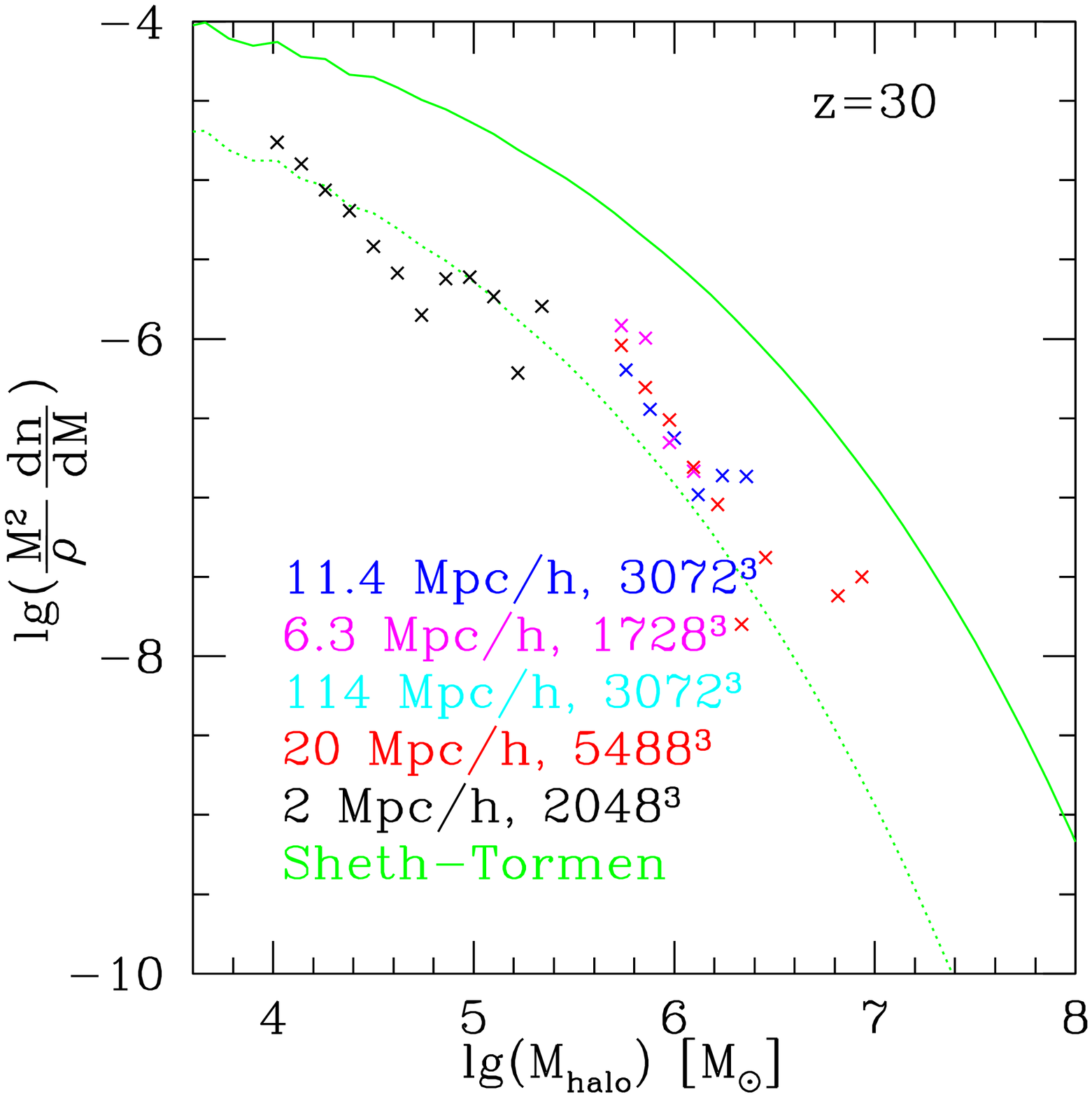}
\includegraphics[width=2.in]{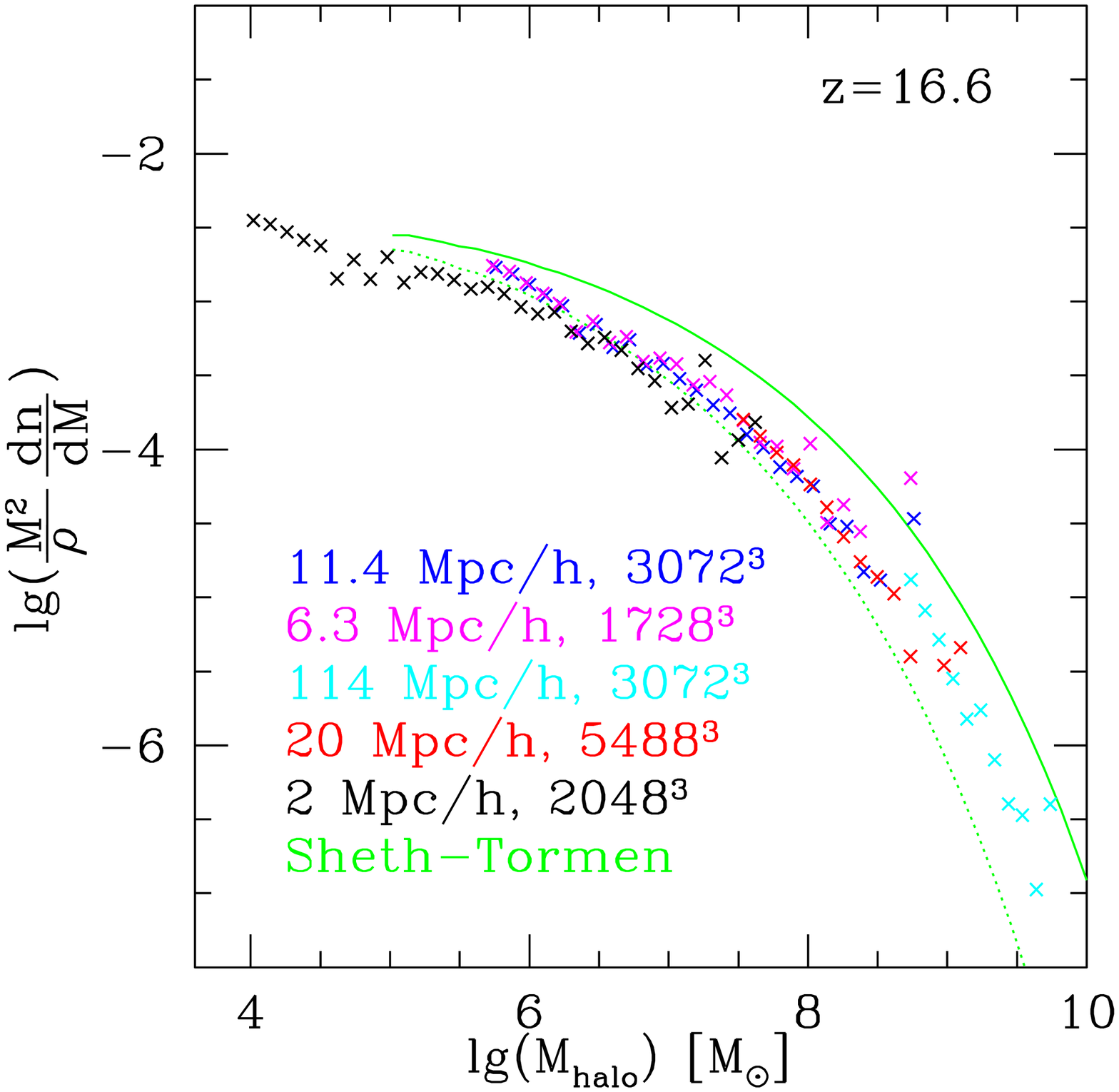}
\includegraphics[width=2.in]{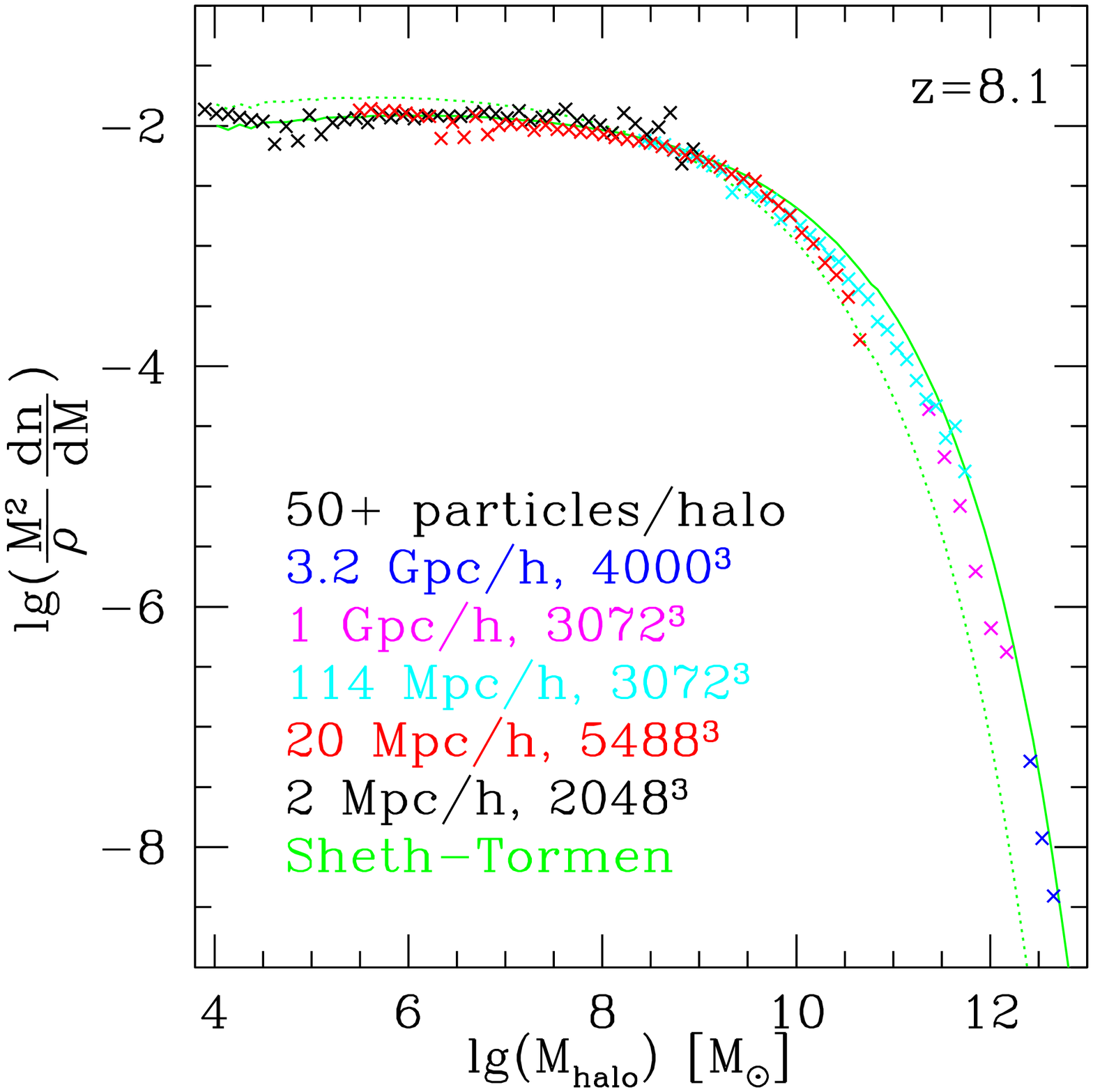}
\vspace{-0.2in}
\caption{Halo multiplicity functions 
at redshifts $z=30$ (left), $z=16.6$ (middle) and $z=8$ (right). 
\label{mult_function_fig}}
\vspace{-0.7cm}
\end{center}
\end{figure}

\vspace{-0.5cm}
\section{Cosmological structure formation}

In Figure~\ref{z9_density_image_fig} we show the matter (green) and halo 
(orange) distributions in a thin slice at redshift $z=8$ from our largest, 
20 Mpc/h box simulation with $5488^3$ particles. The cosmic web is already
well-developed and highly nonlinear at these small scales even at such an 
early time. At this point there are over 110 million resolved dark matter 
halos in the box. The larger, rare halos are strongly clustered, with a
spatial distribution which is highly biased with respect to the underlying
density field, and largely follows the high-density filaments and knots. 
However, there are a fair number of smaller halos (minihaloes) which are 
found in mean and low-density (void) regions. The reason for this is that 
at this time the smallest minihaloes become very common haloes, with
$\nu=\delta_c/\sigma(M)<1$, where $\delta_c\sim1.69$ is the linear 
overdensity at collapse time predicted by the top-hat model and $\sigma(M)$ 
is the density field variance at the appropriate mass scale $M$. A number
of large-scale voids, from a few to $\sim10$~Mpc in size, are found in
our computational volume, as well as a large number of high density peaks.
The density therefore varies very significantly between sub-volumes. For 
example, at $\sim0.5$~Mpc scale the density variation is 1 order of
magnitude even at $z=28$ and reaches $\sim$2 orders of magnitude at $z=8$.

\subsection{Halo mass function}

We locate the collapsed halos on-the-fly, as the code is running, using 
a spherical overdensity halo finder with overdensity of 178. This is 
done by first interpolating the particles onto a gridded density field 
(at resolution twice the number of particles per dimension, as listed in 
Table~\ref{summary_N-body_table}). Local density peaks (with density at 
least 100 times the average) are located and spherical shells are expanded 
around each peak until the threshold overdensity is reached. The resulting 
object is then marked as a halo (objects with less than 20 particles are 
discarded as they cannot be reliably identified). The halo centre position 
is calculated more precisely by quadratic interpolation within the cell and 
the particles within the halo virial radius are identified and then the 
halo properties, e.g. mass, velocity dispersion, center-of-mass, angular 
momentum, radius, etc. are calculated and saved in the halo catalogue.

The resulting halo multiplicity functions, $(M^2/\bar{\rho})(dn/dM)$, at 
$z=30$, 16.6 and 8.1 are shown in Figure~\ref{mult_function_fig}. Here 
we conservatively only include well-resolved halos, with at least 50 
(100) particles at $z=8$ (higher redshifts). The halo mass functions
show significant differences from the widely-used Sheth-Tormen (ST) 
approximation \cite{2002MNRAS.329...61S} (solid green line), particularly 
for rare halos. At $z=8.1$, ST is a reasonably good fit for halos with 
$M_{halo}<10^9M_\odot$ (corresponding roughly to $\nu$ up to a few), but 
over-predicts the abundances of more massive halos by a significant factor 
of up to a few. This is consistent with previous results on the halo mass 
function at high-z \cite{2006MNRAS.369.1625I,2007MNRAS.374....2R,2007ApJ...671.1160L}. At higher redshifts the numerical mass functions do not agree with 
ST by ever larger factors, over-predicting the halo abundances by up to an 
order of magnitude at $z=30$. In fact, at highest redshift, the classic 
Press-Schechter mass function (green dotted line) is a better fit, although 
it somewhat underestimates the halo abundances.


\vspace{-0.1in}
\subsection{Halo bias}
\begin{figure}[ht]
\vspace{-0.4in}
\begin{center}  
\includegraphics[width=2.in]{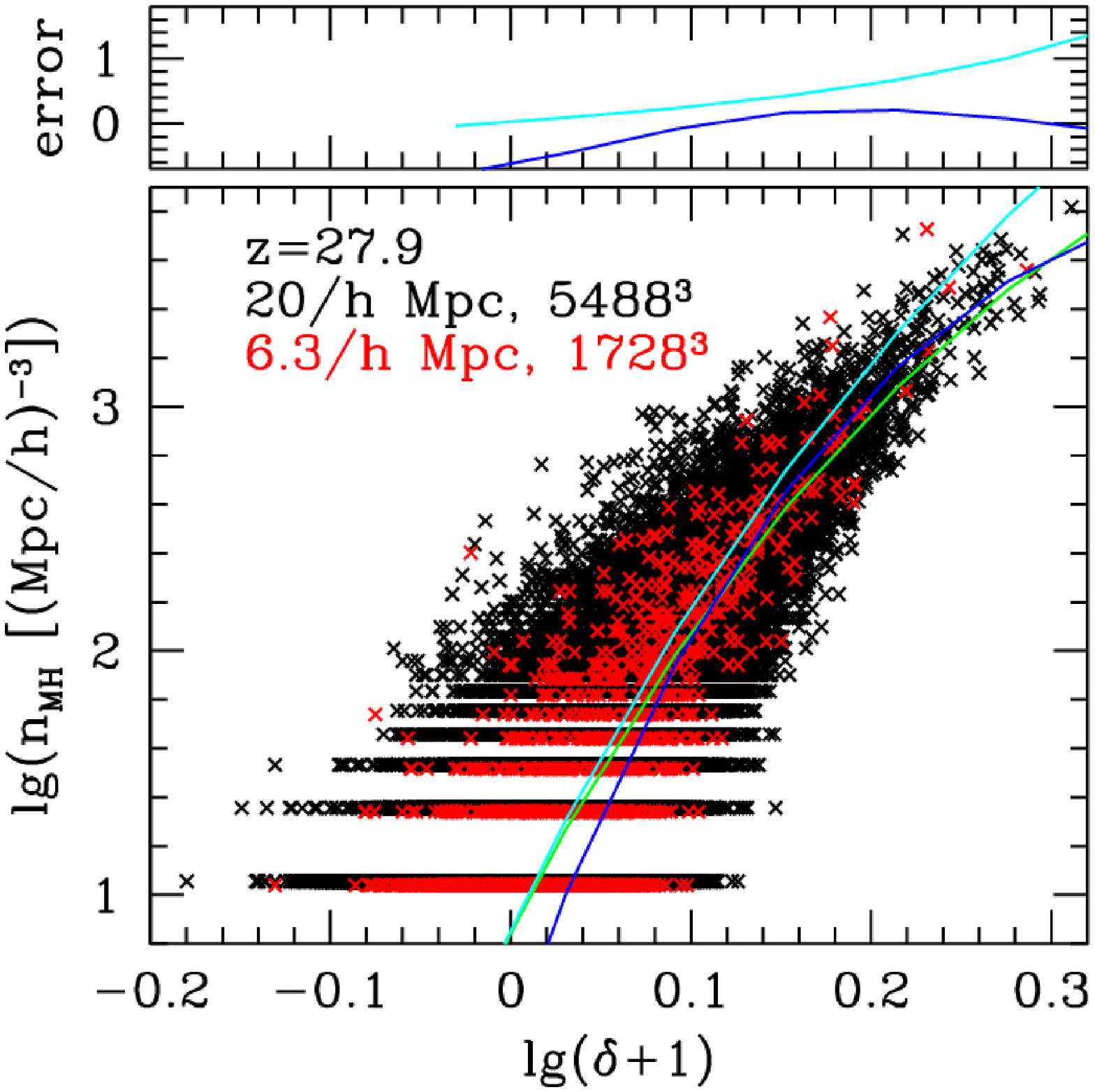}
\includegraphics[width=2.in]{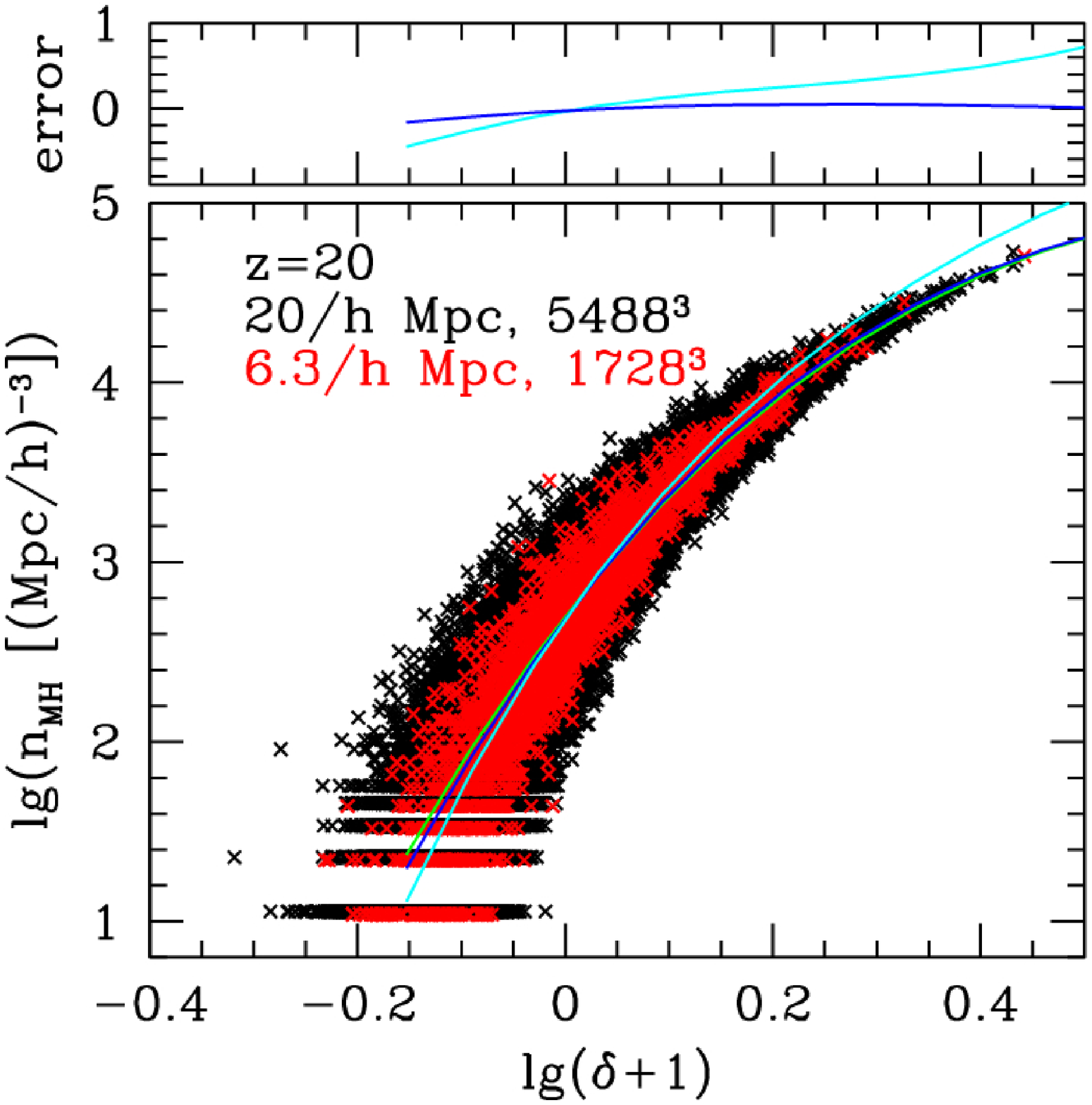}
\includegraphics[width=2.in]{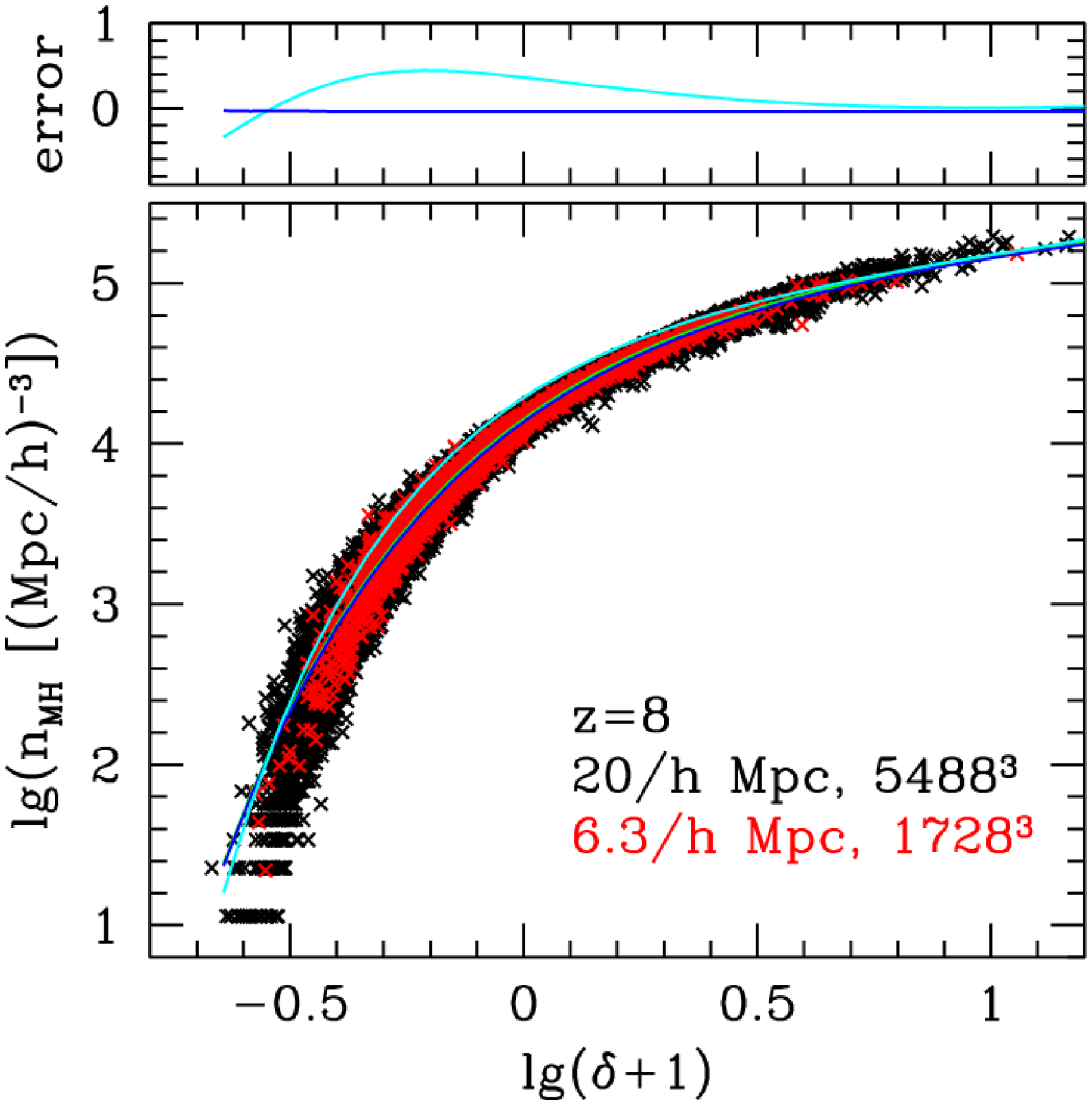}
\vspace{-0.1in}
\caption{Number of minihaloes per unit volume vs. local overdensity, 
$1+\delta$, for sub-volumes of size $440\,{h^{-1}}$kpc at $z=28$ (left),
$z=20$ (middle) and $z=8$ (right). Shown are the results based on 2 
simulations with same resolution, with boxsizes 6.3$\,{h^{-1}}$Mpc (red 
points) and 20$\,{h^{-1}}$Mpc (black points). Also shown are the best 
mean fits for 6.3${h^{-1}}$Mpc (dark blue line) and 
20$\,{h^{-1}}$Mpc (green line) and the extended Press-Schechter (extPS) 
model predictions (light blue line). Top panel shows the extPS (light 
blue) and 6.3$\,{h^{-1}}$Mpc mean (dark blue) in units of the respective 
20$\,{h^{-1}}$Mpc box results.
\label{bias1_fig}}
\vspace{-0.5cm}
\end{center}
\end{figure}

The halo mass function is a strongly nonlinear function of the local density.
Overdense regions behave locally as universe with higher mean density 
producing exponentially more halos. This is directly related to the bias of 
the halo distribution with respect to the underlying density field and is an
important ingredient in many semi-analytical models of structure formation and
reionization. Such models are also used as sub-grid physics in simulations 
when they do not have sufficient resolution to resolve all halos relevant to
the questions being asked. It is therefore very important to have a handle on
the correlation of mass function with local density. To this end, we use data 
from two of our simulations, with box sizes 20~Mpc$/h$, 6.3~Mpc$/h$. 
Scatter plots of the number of minihalos ($M_{halo}<10^8M_\odot$) vs. local 
density in units of the mean are shown in Figure~\ref{bias1_fig}. 
The best-fit mean relations are plotted as well. For comparison we also 
show an analytical bias prescription based on extended
Press-Schechter theory\cite{2004ApJ...609..474B}.  
The first observation is that results are fairly consistent between the 
6.3~Mpc$/h$ and the 20~Mpc$/h$ runs, which have the same spatial and mass 
resolution, but very different volume. Both extremes, high overdensity and 
underdensity, are less well sampled by the smaller box, which is 
especially evident at high redshift, $z=28$, but the best mean fits for each
box agree with each other reasonably well. At later times, $z<20$, they 
become virtually indistinguishable. The analytical model gives 
a relatively good prediction for the correlation at mean density and high 
redshift, and at high density and lower redshift, but can be off by up to
a factor of 2 at other regimes. Finally, we note that there is a significant 
scatter in the halo number - local density relation, particularly at higher 
redshifts. At later times the correlation tightens, although mostly in 
relative terms, because the density variation range increases significantly,
while the absolute value of the scatter remains roughly constant. The 
origins of this scatter and its effects on semi-analytical and sub-grid 
numerical models are currently under investigation.





\vspace{-0.5cm}
\section*{Acknowledgements}

We thank the Texas Advanced Computing Center for providing
HPC resources and excellent support (special thanks to Bill Barth), and 
grants NSF AST 0708176, NASA NNX 07AH09G and NNG04G177G, and Chandra SAO 
TM8-9009X. 

\vspace{-0.4cm}
\section*{References}
\bibliography{../../refs}






\end{document}